\documentclass[aps,
prl,
amsmath,amssymb,
reprint,
superscriptaddress
]{revtex4-2}

\usepackage{graphicx}
\usepackage{dcolumn}
\usepackage{bm}
\usepackage{soul} 
\usepackage[utf8]{inputenc}
\usepackage[T1]{fontenc}
\usepackage{mathptmx}
\usepackage{etoolbox}
\usepackage{color}

\newcommand{\BXf}{B_{\mbox{\tiny Xf}}}
\newcommand{\kB}{k_{\mbox{\tiny B}}}

\newcommand{\TN}{T_{\mbox{\tiny N}}}

\newcommand{\muEu}{\mu_{\mbox{\tiny Eu}}}

\newcommand{\Bint}{B_{\mbox{\scriptsize int}}}

\newcommand{\DeltaThetaFSAT}{\Delta\theta_F^{\mbox{\tiny SAT}}}
\newcommand{\DeltaThetaF}{\Delta\theta_F}

\newcommand{\muPol}{\mu_{\mbox{\tiny Pol}}}

\newcommand{\muPolFG}{\mu_{\mbox{\tiny Pol}}^{\mbox{\tiny FG}}}

\begin{document}

\title{
Direct optical probing of ultrafast spin dynamics in a magnetic semiconductor}
\author{S. C. P. van Kooten}
\affiliation{Laboratório de Magneto-Óptica - LMO, Instituto de Física, Universidade de São Paulo, 05508-090 São Paulo, Brazil}
\author{G. Springholz}
\affiliation{Institut f\"ur Halbleiter und Festk\"orperphysik, Johannes Kepler Universit\"at Linz, 4040 Linz, Austria}
\author{A. B. Henriques}
\affiliation{Laboratório de Magneto-Óptica - LMO, Instituto de Física, Universidade de São Paulo, 05508-090 São Paulo, Brazil}

\date{\today}

\begin{abstract}
We uncovered the spin dynamics involved in the birth and growth of giant spin polarons in a magnetic semiconductor. For this purpose, we developed a new measurement technique, which provides direct access to the spin dynamics, irrespective of phonons and carriers involved in the process. Moreover, we solved the Landau-Liftshitz equation in the specific scenario of spin polarons, which fits our data excellently, and demonstrates that the spin polaron growth slows down dramatically when the sample is cooled in the paramagnetic phase. Finally, temperature dependent Monte Carlo simulations were performed, which are in excellent agreement with the observed slowdown, which demonstrates that fluctuations in the Weiss field play a decisive role in spin coherence generation induced by light in magnetic materials. These results offer a new tool and new insight for spin dynamics investigations.
\end{abstract}

\maketitle

The control of the magnetic state of matter with light is a topic of vast current technological and scientific interest \cite{kirilyukRMP,naturephys2018,nature2019,KimelMar2019}.
Understanding the physics behind light-induced magnetization is essential to develop more efficient magneto-optical devices. Previous magnetization dynamics investigations focused mainly on demagnetization driven by the heat generated by the pump pulse \cite{BeaurepaireMay1996,stanciuPRL2007,koopmansNAT2010,kimelAPL2022}, as well as on some limited non-thermal mechanisms based on e.g. photoinduced changes in magnetic anisotropy \cite{DuongSep2004,HansteenJul2005,StupakiewiczFeb2017} and coherent stimulated Raman scattering \cite{KimelJun2005}.

Recently, we reported the discovery of efficient non-thermal light-induced magnetization generation in intrinsic magnetic semiconductors, based on the photoexcitation of very large spin polarons (SPs), i.e. bound electron-hole pairs forming excitons, within which thousands of magnetic atoms are forced into ferromagnetic alignment due to the band-lattice exchange interaction \cite{prl2018,prb16Rapid,prb2014,prb09}.
Here we extract for the first time the spin dynamics involved in the birth and growth giant SPs. For this purpose, we developed a new technique, which extracts spin dynamics directly from measurements, irrespective of the phonon and carrier subsystems. Until now, extracting spin dynamics from transient optical measurements demanded a model involving all three interacting subsystems perturbed by the incident light: carriers, phonons, and spins. Because our technique provides direct access to spin dynamics, irrespective of carriers and phonons, it uncovers the spin dynamics unequivocally, and opens a new avenue for the investigation of light-induced magnetism. Moreover, the experimental results were analyzed in the frame of the Landau-Lifshitz equation, which was solved in the specific scenario of SP generation, yielding the SP growth rate and fully-grown size. Additionally, we demonstrate the crucial role played by Weiss field fluctuations in the spin dynamics. Due to the high interest in the control of spin phenomena using light, our results will attract the attention of a broad audience.

The EuSe samples were grown by molecular beam epitaxy onto (111) BaF$_2$ substrates at the Johannes Kepler Universit\"at Linz. Because of the almost perfect lattice constant matching ($a=6.191~{\mbox{\AA}}$ and $a=6.196~{\mbox{\AA}}$ for EuSe and BaF$_2$, respectively), nearly unstrained bulklike EuSe reference layers with $\mu$m thickness were obtained by direct growth on BaF$_2$. Results presented here were obtained on an EuSe epilayer of thickness 0.34~$\mu$m.
The time-resolved photoinduced Faraday rotation (TRFR) was measured at the LMO of the University of S\~ao Paulo, using a two-color pump-probe technique. We used 100~fs light pulses, tuned to a repetition rate of 25~kHz, produced  by a cavity-dumped Ti:Sapphire mode-locked laser coupled to a second harmonic generator. The pump photon energy was 3.1~eV, which is above the EuSe band gap of 2.0~eV, therefore exciting electron-hole pairs that form supergiant SPs \cite{prl2018,jap2022}, whereas the probe photon energy was 1.55~eV, which is in the transparency range of EuSe, but sufficiently close for efficient Faraday rotation detection of magnetization \cite{jap2019}.
The absorption coefficient for EuSe at the 3.1~eV excitation energy is greater than 20~$\mu$m$^{-1}$ \cite{prb05}, hence the penetration depth of the excitation light is only 100~nm or less.
The delay between the pump and probe pulses was controlled by a 0.6~m delay line. The Faraday rotation angle of the probe pulse was measured using autobalanced detection  with a resolution better than 0.1~$\mu$rad. The image of the excitation spot on the sample had a diameter of 150~$\mu$m, about twice the diameter of the probe spot. All measurements were performed using an optical cryostat containing a superconducting coil, applied in the Faraday geometry. Measurements were taken above the N\'eel temperature, when EuSe is in the paramagnetic phase and the equilibrium magnetization is zero. A remarkable feature of the magnetization generated by SPs is that it is triggered by very low intensity light, which offers the advantage of investigating spin dynamics under minimal thermal perturbation. Our investigation was performed using extremely low fluence light pulses, of about 1--10~$\mu$J/cm$^2$, which is several orders of magnitude less than employed in the investigations by other authors.

\begin{figure}
\includegraphics[angle=0,width=90mm]{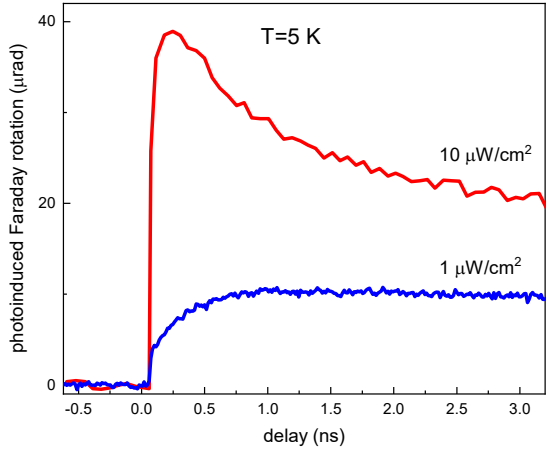}
\caption{Time-resolved Faraday rotation in EuSe for different fluences at T = 5~K and a magnetic field of 60~mT.}
\label{fig:delayScan}
\end{figure}

Figure \ref{fig:delayScan} shows a delay scan of the TRFR for fluences of 1 and 10~$\mu$J/cm$^2$. As demonstrated in Ref.~\onlinecite{prl2018}, the Faraday rotation signal is due to photoexcited supergiant spin polarons (SPs). The time interval between pulses, 40~$\mu$s, is much longer than the SP lifetime of about 1~$\mu$s \cite{prl2018}, therefore allowing for the spin system to return to equilibrium before the arrival of the next pulse.
In EuSe, SPs can be photogenerated at very high densites to completely fill the illuminated volume, opening the perspective of achieving a large magnetization with low intensity light.
The population of SPs generated at $t=0$ remains effectively constant over the positive time delay of a few nanoseconds shown in Fig.~\ref{fig:delayScan}, because it is orders of magnitude shorter than the SP lifetime.

For 1~$\mu$J/cm$^2$ pulses, the shape of the TRFR displays exponential-like growth after the arrival of the pump pulse at $t=0$. As pulse fluence increases, a cusp begins to form immediately after $t=0$, becoming very prominent for 10~$\mu$J/cm$^2$ pulses. The emergence of a cusp when the pump energy increases can be qualitatively explained by the
fact that the pump pulse perturbs three subsystems - charge carriers, phonons, and lattice spins. The pump light generates electrons in excited conduction band states; as the electrons relax their energy and form SPs, heat is transferred to the phonon and spin baths, increasing the lattice temperature, which reduces the magnetic susceptibility, and consequently the photoinduced magnetization signal decreases, producing the cusp. As the excitation fluence is increased, more heat is generated, leading to a more prominent cusp in the delay scan. This demonstrates that the delay scan shown in Fig.~\ref{fig:delayScan} reflects simultaneous changes in the interacting subsystems, hence extraction of spin dynamics
from such a delay scan mandates the assumption of an underlying model involving all three subsystems, which is the approach used by many researchers \cite{BeaurepaireMay1996,koopmansNAT2010,BattiatoJul2010}.

In this report, we circumvent the requirement of such a three-subsystem model, and extract the time evolution of the spin subsystem independently of the carrier and phonon subsystems. This is achieved by measuring the TRFR as a function of magnetic field at constant time delays, from which we then extract spin dynamics associated with SP generation.
At each time delay the state of the carrier, phonon and spin subsystems remain frozen, and TRFR magnetic field scans reveal the state of the spin subsystem at that particular instant of time.
The magnetic field scans at fixed delays covered only a fraction of a Tesla, which is sufficient to polarize the SPs, in view of their very large magnetic moment, but has virtually no direct effect on individual Eu spin alignment.

\begin{figure}
\includegraphics[angle=0,width=90mm]{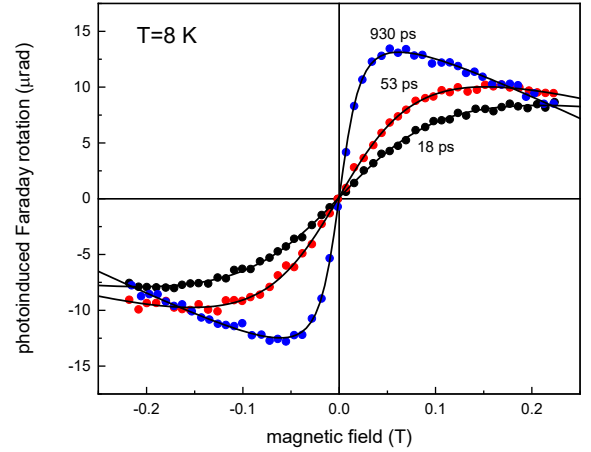}
\caption{Field scans at various fixed time delays.}
\label{fig:magscan}
\end{figure}

To minimize heating effects, we used a pump fluence of 1~$\mu$J/cm$^2$ for the remainder of this work. Figure~\ref{fig:magscan} shows typical magnetic field scans, taken for various time delays at $T=8$~K.

For the low excitation levels used in this work, the photoinduced SPs are non-interacting, forming a superparamagnetic system \cite{prb16Rapid,prb17EuTe}, therefore the photoinduced Faraday rotation angle, $\DeltaThetaF$, which is proportional to the magnetization \cite{jap2019}, is described by a Langevin function $L(x)$:
\begin{equation}
\DeltaThetaF=\DeltaThetaFSAT\, L\left(\frac{\muPol\Bint}{\kB T}\right)
+\mbox{\rm linear term}.
\label{eq:Langevin}
\end{equation}
where $\DeltaThetaFSAT$ is the Faraday rotation due to a fully polarized SP gas, $\muPol$ is the SP magnetic moment, $\Bint$ is the internal magnetic field, and $T$ is the temperature of the lattice within the SP.
The linear term in \eqref{eq:Langevin}
describes the heating caused by the pump pulse, as described in detail in Ref.~\onlinecite{prb17EuTe}.
Solid lines in Fig.~\ref{fig:magscan} represent the theoretical fitting of the experimental data, using equation \eqref{eq:Langevin}. It can be seen that
formula \eqref{eq:Langevin} fits perfectly to the experimental data for all delays. From the fitted curves we extract, at each delay, the amplitude of the magnetization curve, $\DeltaThetaFSAT$, the magnitude of the SP magnetic moment, $\muPol$, and the slope of the linear term. The linear term has a negative slope, exactly as expected when the semiconductor is in the paramagnetic phase \cite{prb17EuTe}.
The absolute value of the slope of the linear term also grows with delay, indicating warming up of the SP volume, which occurs as the extra energy of excited carriers is transferred to the lattice.
Here we shall focus on $\muPol$, whose time evolution contains all necessary information about
the spin dynamics, which is the subject of the present investigation.

\begin{figure}
\includegraphics[angle=0,width=90mm]{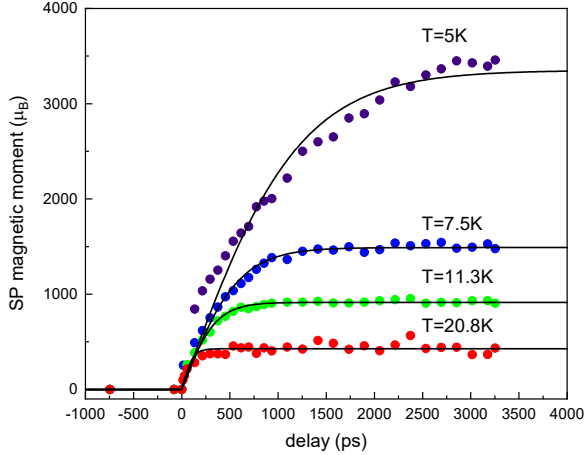}
\caption{Spin polaron magnetic moment as a function of delay for several temperatures. Dots represented experimental data, whereas the solid lines are the results of fits using equation \eqref{eq:muPOLvsTime}. The fits yield the SP growth rate, $T_1$, and magnetic moment of the fully grown quasi-stationary SPs, $\muPolFG$.}
\label{fig:muPolVsDelay}
\end{figure}

\begin{figure}
\includegraphics[angle=0,width=90mm]{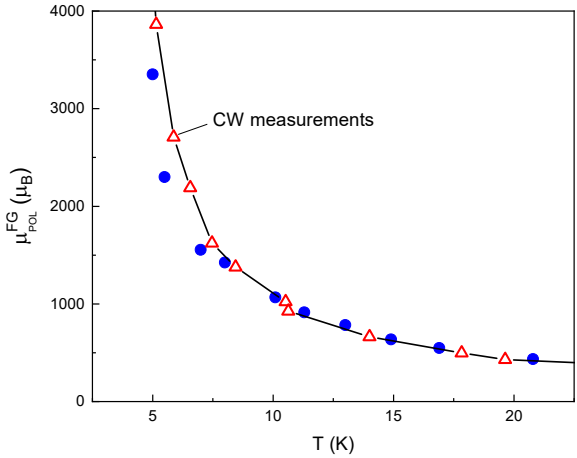}
\caption{Time dependence of the SP magnetic moment measured by TRFR at saturation delays (circles) and by CW measurements (triangles connected by a straight line).}
\label{fig:muPolVsT}
\end{figure}

Figure~\ref{fig:muPolVsDelay} shows the magnitude of the SP magnetic moment, in Bohr magneton ($\mu_B$) units, as a function of delay, measured for several temperatures.
We observe that $\muPol$ grows over time, and at each temperature it tends to a constant value, $\muPolFG$, which describes the
magnetic moment of a fully grown SP.

In Fig.~\ref{fig:muPolVsT} we compare the saturation moment $\muPolFG$ with measurements done using continuous wave (CW) laser excitation, taken from Ref.~\onlinecite{prl2018}.
Under CW pumping, a constant population of fully grown SPs is maintained \cite{prb16Rapid,prl2018}.
Figure~\ref{fig:muPolVsT} reveals that the saturation moment, $\muPolFG$, found from the TRFR measuremensts at long delays, coincides perfectly with CW data for the magnetic moment of SPs found from CW  measurements. This proves that the time dependence of $\muPol$ extracted from the magnetization curves at fixed delays, shown in Fig.~\ref{fig:muPolVsDelay}, is a direct observation of the birth and growth of the supergiant SP in EuSe, and therefore the time evolution of $\muPol$ as a function of time, at a fixed temperature, is directly connected to the spin dynamics of the Eu atoms in the crystal.

The mathematical connection between spin dynamics and the growth of the magnetic moment of SPs is provided by the Landau-Lifshitz equation for the expectation value of a lattice spin vector, $\bm S$,
\begin{equation}
\frac{d{\bm S}}{dt}={\bm\omega}\times{\bm S}+\frac{\alpha}{S}\,{\bm S}\times\left({\bm\omega}\times{\bm S}\right),
\label{eq:LL}
\end{equation}
where $\bm\omega=g\mu_B\bm B$, $g=2$ is the gyromagnetic factor of an europium atom in the lattice, $\bm B$ is the effective magnetic field acting on the Eu atom, and $\alpha$ is a dimensionless coefficient associated with the directional relaxation of the spin vector
$\bm S$. The first term in \eqref{eq:LL} describes the precession of $\bm S$ around $\bm B$, and follows from the Schr\"odinger equation with Zeeman interaction. The second term is phenomenological, corresponding to the directional relaxation of ${\bm S}$ towards $\bm B$, with conservation of the absolute value of $\bm S$ (see, for instance, Ref.~\onlinecite{gurevich}).

If $\bm B$ is constant, then the solution of \eqref{eq:LL}, averaged over random initial orientations that characterize an Eu spin in the paramagnetic phase, is given by
\begin{equation}
\overline{S_z}=S\,\tanh\left(\frac{t}{2T_1}\right),\,\,
\overline{S_x}=\overline{S_y}=0,
\label{eq:Sparamag}
\end{equation}
where $T_1=1/\alpha\omega$ is the characteristic time for the directional relaxation of the Eu spins towards $\bm B$, and $z$ is the direction of $\bm B$.

When an electron is photoexcited from the localized $4f^7\left(^8S_{7/2}\right)$ valence state of an Eu atom by an absorbed photon, the electron is promoted to a conduction band state, which overlaps with many lattice sites, and therefore it suddenly generates an exchange field, $\BXf$, acting on the surrounding Eu atoms
\cite{prb2014,prb08,prb09,prl2009,jpc07}. The average value within the SP sphere, $\overline\BXf$, is about 0.7~Tesla for EuSe \cite{apl2011,prl2018}. It is this field that polarizes the lattice spins, forming an SP. Therefore the exchange field entering \eqref{eq:LL} can be approximated by $\overline{\BXf}$. From \eqref{eq:Sparamag}, the growth of the SP magnetic moment is then given by
\begin{equation}
\muPol(t)=\muPolFG\,\tanh\left(\frac{t}{2 T_1}\right),
\label{eq:muPOLvsTime}
\end{equation}
which indicates that $T_1$ is also the characteristic time for SP growth.

The solid lines in Fig.~\ref{fig:muPolVsDelay} indicate a fit of \eqref{eq:muPOLvsTime} to the experimental data. A very good fit is found, although a slight deviation is observed at $T=5$~K, probably because it is too close to the EuSe N\'eel temperature ($\TN$=4.8~K \cite{prl2018}), leading to some correlations between spins and invalidating the assumption of ideal paramagnetic order.

\begin{figure}
\includegraphics[angle=0,width=90mm]{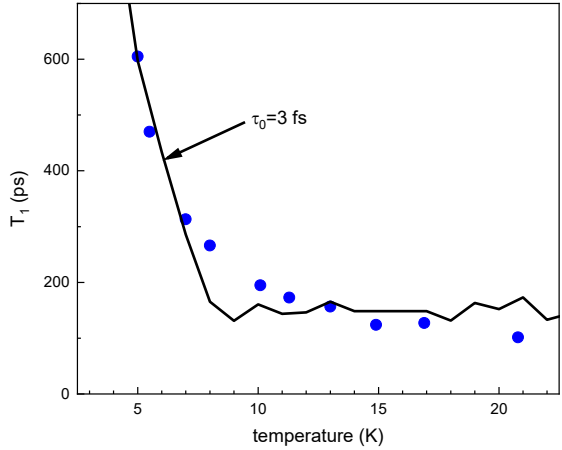}
\caption{Dots represent the SP characteristic growth time, $T_1$, as a function of temperature, obtained from TRFR measurements. The solid line represents the result of Monte Carlo simulations, from which the Weiss field fluctuation rate is found to be $\tau_0=3$~fs.}
\label{fig:tauVsT}
\end{figure}

Figure~\ref{fig:tauVsT} displays the extracted rise time $T_1$ as a function of temperature, showing that the SP growth slows down dramatically as the sample is cooled towards the N\'eel temperature. We attribute this slowdown to fluctuations of the Weiss exchange field, as will be shown henceforward.

Directional spin relaxation towards $\bm B$
can be brought about by a transverse oscillating magnetic field,
as in nuclear magnetic resonance, in which case the relaxation rate is proportional to
the squared amplitude of the oscillating magnetic field
(see, for instance, Ref.~\onlinecite{Kittel}). In the case of a magnetic material, such as EuSe, fluctuations in Weiss magnetic field can similarly assist the directional relaxation of $\bm S$ towards $\bm B$. Since the amplitude of fluctuations decreases as temperature approaches $\TN$, the relaxation rate assisted by Weiss field fluctuations also decreases, which explains qualitatively the slowdown of the SP growth seen in Fig.~\ref{fig:tauVsT}.

For a quantitative analysis, we shall use the formula connecting the relaxation rate, $T_1$, and Weiss field fluctuations, $\delta B_\perp$, described in the book by A. P. Guimar\~aes (Ref. \onlinecite{guimaraes}):
\begin{equation}
T_1=\frac{1+\omega_0^2\tau_0^2}{\gamma^2\tau_0\delta B_\perp^2},
\label{eq:T1}
\end{equation}
where $\gamma=\muEu/\hbar$, $\muEu=g\mu_B S$ is the magnetic moment of an Eu atom, $g=2$, $S=7/2$, $\omega_0=\gamma\overline\BXf$, and $\tau_0$ is the typical interval of time for which the Weiss field remains constant. We calculated the amplitude of fluctuation of the Weiss field, $\delta B_\perp$ as a function of temperature using Monte Carlo calculations, fully described in Refs.~\onlinecite{apl2020,prb2021,jap2022}, and used $\tau_0$ as an adjustable parameter in \eqref{eq:T1} to fit the experimental values of $T_1$. As depicted in Fig.~\ref{fig:tauVsT}, a good fit of the data is found assuming a temperature-independent value $\tau_0=3$~fs, which is of the same order of magnitude as the spin correlation time found in other magnetic materials \cite{SpinCorrelTimeNatcom2018}. As temperature increases, the calculated $T_1$ flattens compared to a slight, continued decrease of the experimental curve, which suggests that $\tau_0$ decreases.

In conclusion, we demonstrated a novel method to extract spin dynamics, which does not require modeling of the phonon and carrier subsystems, which obviously makes the spin dynamics extracted more accurate. We combined our new method with our solution of the Landau-Lifshitz equation to describe spin dynamics, and measured the growth rate and fully-grown size of supergiant spin polarons in EuSe. We demonstrated that in the paramagnetic phase the spin dynamics depends dramatically on temperature. Using Monte Carlo simulations we demonstrate the vital role played by Weiss field fluctuations. The connection between Weiss field fluctuation and spin dynamics discovered by us offers new into the optical control of spin coherence. These results offer a new efficient tool and new insight for spin dynamics investigations in magnetic materials.

A. B. H. acknowledges support by CNPq (Grant Nos. 420531/
2018-1 and 303757/2018-3) and FAPESP (Grant No. 2020/15570-0). S. C. P. K. acknowledges support by CAPES (doctoral fellowship).

\providecommand{\noopsort}[1]{}\providecommand{\singleletter}[1]{#1}%

\end{document}